# *ab initio* informed inelastic neutron scattering for time-resolved local dynamics in molten MgCl$_2$

Shubhojit Banerjee,[1*] Rajni Chahal,[3] Alexander S. Ivanov,[3] Santanu Roy,[3] Vyacheslav S. Bryantsev,[3] Yuya Shinohara,[2] Stephen T Lam[1*]

[1] University of Massachusetts Lowell, 1 University Ave, Lowell, MA, 01854

[2] Materials Science and Technology Division, Oak Ridge National Laboratory, Oak Ridge, Tennessee 37831, United States

[3] Chemical Sciences Division, Oak Ridge National Laboratory, Oak Ridge, Tennessee 37831, United States

AUTHOR INFORMATION

Corresponding Authors:

*Shubhojit Banerjee: shubhojit_banerjee@student.uml.edu

* Stephen T Lam: stephen_Lam@uml.edu

**Abstract**

Ion dynamics that drive the transport and thermophysical properties of molten salts are poorly understood due to challenges in precisely quantifying the spatial and temporal fluctuations of specific ions in highly disordered systems. While the Van Hove correlation function (VHF) obtained from inelastic neutron scattering (INS) probes these dynamics directly, its interpretation is limited by the inherent species-averaging of experiments, which obscures analysis of key ion transport and solvation mechanisms. Here, *ab initio* molecular dynamics (AIMD) is used to model the VHF, unravel its partial contributions, and elucidate its underlying ionic transport mechanisms. Slow decorrelation is revealed for oppositely charged ions ($Mg^{2+}$ and $Cl^-$) caused by ion exchange across the solvation shell between adjoining ionocovalent complexes. Furthermore, transport coefficients are accurately recovered and connections between macroscopic properties and ion dynamics are revealed. This study demonstrates the potential of *ab initio*-informed VHF to resolve long-standing challenges in uncovering relationships between picosecond-scale ion dynamics, mechanisms, and emergent physical properties of molten salts.

**TOC GRAPHIC**:

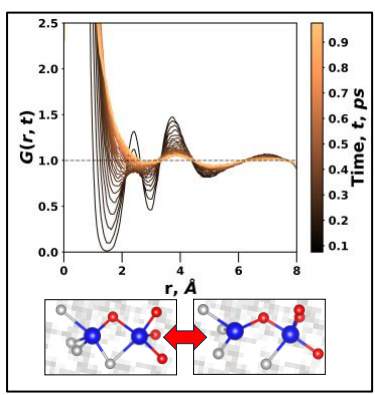

**KEYWORDS**: Molten salt, $MgCl_2$, liquid dynamics, Van Hove correlation function, PMF, inelastic neutron scattering, Cage correlation function, Diffusion coefficient.

Molten salts are promising candidate heat transfer fluids for solar thermal energy storage and advanced nuclear reactors due to their high boiling point, low vapor pressure, and favorable heat transfer properties.[1–5] In particular, $MgCl_2$-based salts are commonly used across various advanced reactor and fuel processing applications due to their low neutron absorption, low melting points, and their ability to dissolve fuel constituents.[4–7] Recent *ab initio* simulations have shown that ion-pair dissociation processes occur at the picosecond timescale. Such dynamics are known to play a fundamental role in determining various thermodynamic and transport properties of liquid metals and molten salts, namely activity coefficients, diffusivity, and viscosity.[8–11] Yet, most prior experimental studies involving X-ray, neutron, and Raman spectroscopy[12,13] reveal only the equilibrium ensemble-averaged local structures via the pair distribution function. In contrast, a recent study[14] showed that inelastic neutron scattering could potentially be used to ascertain the kinetics of ion exchange across the solvation shell (SS) in $MgCl_2$ molten salt using the Van Hove correlation function (VHF). However, a significant challenge lies in interpreting and extracting precise physical behavior directly from these measurements due to the information being hidden in the total spectrum, which combines correlation functions of all ionic pairs (*i.e.,* cation-cation, cation-anion, and anion-anion). As such, these measurements must be further deconvoluted to describe and completely understand element-specific ion-ion dynamics.

Here, we demonstrate that predictive molecular simulations can play a vital role in assisting, enhancing, and validating the interpretation of the VHF spectra to reveal element-specific dynamics.[15] Using *ab initio* molecular dynamics (AIMD), individual ion interactions and dynamics are simulated explicitly, and the element-specific contributions towards the ensemble-averaged VHF are accessed to deconvolute the experimental spectra. In this letter, we study the sub-picosecond dynamic structure of molten $MgCl_2$ at 990 K, 1148 K, and 1600 K. AIMD simulations were conducted to replicate and extend the thermodynamic conditions of the previous experiments, effectively maximizing the information that can be extracted from expensive and time-intensive INS studies.

Simulations are first validated against the average local structure determined in previous experiments. A detailed setup of all simulations is described in the Methods Section. First, the local structure is examined through the ensemble-averaged radial distribution functions (RDF) for Mg-Cl, Mg-Mg, and Cl-Cl pairs. The RDFs are shown in Figure S1(a)-(c), in which simulated peak distance $r_{Mg-Cl}^{sim} = 2.38 \pm 0.05$ Å (the errors are reported as the standard errors computed from RDF data with varying bin size) is nearly identical to the experimental peak distance $r_{Mg-Cl}^{expt} = 2.42 \pm 0.03$ Å at 998 K.[16] Integrating the RDF up to the first minima (*i.e.,* 3.45 Å, 5.1 Å, and 5.2 Å for Mg-Cl, Cl-Cl, and Mg-Mg, respectively), the simulated coordination numbers are found to be $n_{Cl-Cl}^{sim} = 11.6$, $n_{Mg-Cl}^{sim} = 4.4$, and $n_{Mg-Mg}^{sim} = 4.9$. These coordination values are in exact agreement (within experimental uncertainty) with past neutron scattering experiments at a similar temperature (~998K), which found $n_{Cl-Cl}^{expt} = 12 \pm 1$, $n_{Mg-Cl}^{expt} = 4.3 \pm 0.3$, and $n_{Mg-Mg}^{expt} = 5 \pm 1$.[16] These results demonstrate that simulations accurately predict average local coordination environment in terms of bond lengths and coordination numbers.

Beyond average ion coordination, complexes and their connectivity are relevant to dissociation mechanisms. This is examined at 990 K, 1148 K, and 1600 K with RDF and angular distribution functions (ADFs) shown in Figure S1(a)-(d). At all temperatures, predominantly four, five, and six coordination states of the Mg were observed as $[MgCl_4]^{2-}$, $[MgCl_5]^{3-}$, and $[MgCl_6]^{4-}$. Moreover, the main peak in Cl-Mg-Cl (upper solid lines in Figure S1(d)) reveals the presence of tetrahedral $[MgCl_4]^{2-}$ complexes, while two distinct peaks in the Mg-Cl-Mg ADF (lower dotted lines) reveal 'corner-sharing' and 'edge-sharing' of those complexes, respectively. These results are all consistent with the analysis from Raman spectra[12] and other theoretical studies.[3,17,18] Overall, these results provide validation of AIMD for accurately predicting salt structures in terms of co-existing coordination, connected molecular complexes, and their topology.

Using AIMD protocols validated on ensemble-averaged structures, the VHF ($G(r,t)$) was simulated to study the dynamic ion-ion structure (Figure 1), including the partial element-specific VHF ($G_{\alpha\beta}$)[19] shown in Equation 1.[15] The total VHF is constructed by a linear combination of the partial VHFs as shown in Equation 2.

$$G_{\alpha\beta}(r,t) = \frac{V}{4\pi N_\alpha N_\beta r^2} \sum_{i \in \{\alpha\}} \sum_{j \in \{\beta\}} \delta(r - |r_i(0) - r_j(t)|) \quad (1)$$

$$G(r,t) = \left(\frac{1}{B}\right)\left(2 * \chi_\beta \chi_\alpha \overline{b_\alpha}\,\overline{b_\beta}\, G_{\alpha\beta}(r,t) + \chi_\alpha^2 \overline{b_\alpha^2}\, G_{\alpha\alpha}(r,t) + \chi_\beta^2 \overline{b_\beta^2}\, G_{\beta\beta}(r,t)\right) \quad (2)$$

In Equation 1, $N_\alpha$ and $N_\beta$ are the number of atoms of species $\alpha$ and species $\beta$, respectively. In equation 2, the $\chi_\alpha$ and $\chi_\beta$ are the atomic fractions of type $\beta$ and type $\alpha$, respectively, $\overline{b_\alpha}$ and $\overline{b_\beta}$ represent neutron scattering length taken from the NIST database, $\delta(r)$ is the Kronecker delta function, and $B$ is the sum of partial contribution coefficients ($2\chi_\beta \chi_\alpha \overline{b_\alpha}\,\overline{b_\beta}\, G_{\alpha\beta} + \chi_\alpha^2 \overline{b_\alpha^2} + \chi_\beta^2 \overline{b_\beta^2}$).

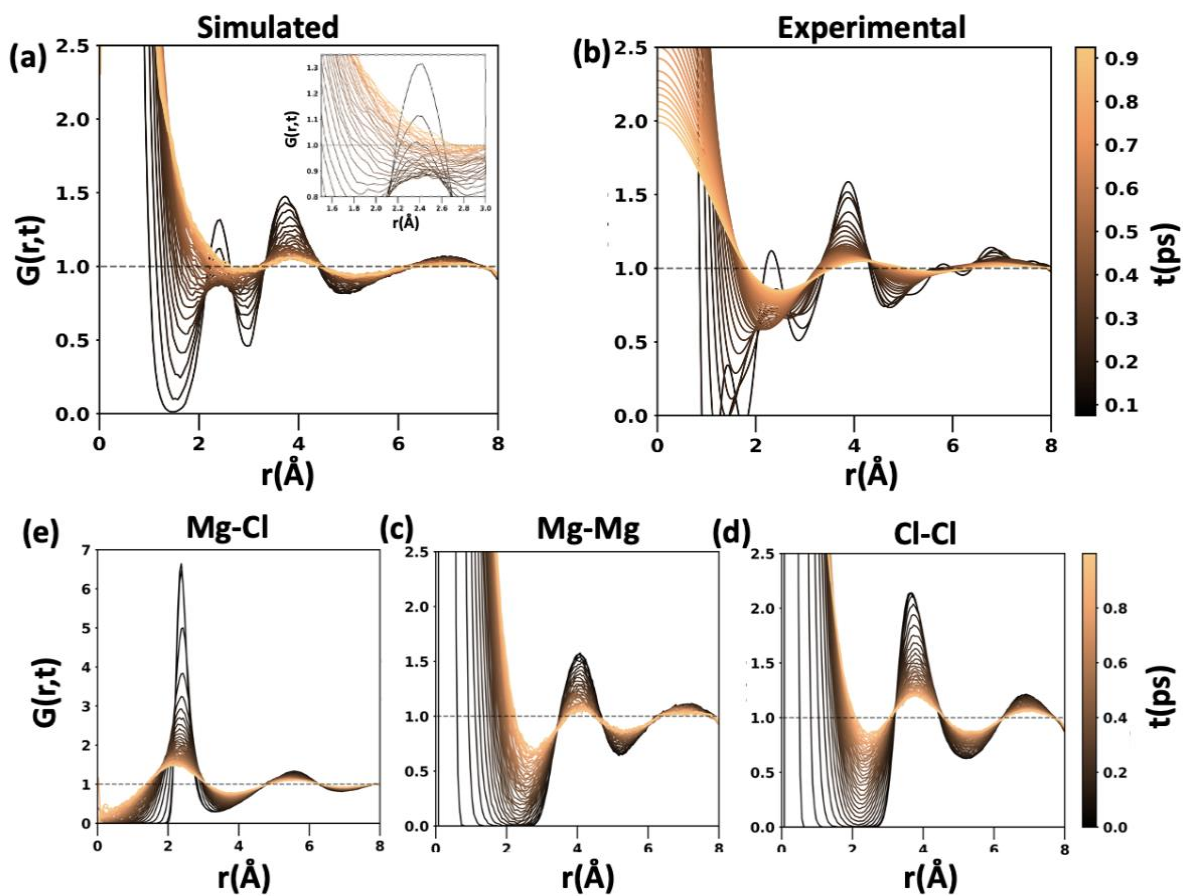

**Figure 1**. Local ion dynamics in MgCl$_2$ at 1148 K given by total VHF obtained from AIMD simulations. (a) Simulated VHF spectra (in inset zoom around the 2.4 Å peak is shown) and (b) INS experimental spectra[14] starting from $t = 0.075$ ps with $\Delta t = 0.025$ ps. Partial VHFs for (c) Mg-Cl, Mg-Mg (d), and (e) Cl-Cl (e) starting from $t = 0$ with $\Delta t = 0.025$ ps.

In Figures 1(a) and 1(b), the simulated VHFs agree with those obtained from INS in the correlation peak distances, intensities, and correlation decay rate; detailed comparisons are shown in Figure S2. Specifically, simulated and experimental total VHFs have similar spectral features with two positive correlation peaks ($G(r,t) > 1$) at around 2.4 Å and 3.8 Å. By comparison with the partial $G_{\text{Mg}-\text{Cl}}$ in Figure 1(c), the first peak in the total VHF at 2.4 Å can be attributed to coordination between oppositely charged ions, Mg$^{2+}$ and Cl$^-$. Meanwhile, the peak around 3.8 Å in the total VHF spectra can be attributed to the correlation of similar charge ions (Mg-Mg and Cl-Cl) based on comparison to $G_{\text{Mg}-\text{Mg}}$ and $G_{\text{Cl}-\text{Cl}}$ shown in Figure 1(d) and 1(e), respectively. In terms of the peak intensities in the total VHF (main experimental observable), the positive correlation between opposite-charge ions (peak at 2.4 Å) appears to decay rapidly and rise again, while the correlation between same-charge ions (peak at 3.8 Å) monotonically decays at a significantly slower rate. Without analysis of element-specific contributions, this cannot be readily explained by known interactions between charged particles.[14]

To advance the understanding of the correlated motion of ions, element-specific VHFs were calculated following the Faber–Ziman formalism of partial structure factor, as shown in Figures 1(c)-(e). The VHFs can be further decomposed into the distinct part $G_{\alpha\beta}^d(r,t)$ and self-part $G_s(r,t)$, which are shown in Equation 3a and 3b, respectively.

$$G_{\alpha\beta}^d(r,t) = \frac{V}{4\pi N_\alpha N_\beta r^2} \sum_{i\in\{\alpha\}} \sum_{j\in\{\beta\}} \delta(r - |r_i(0) - r_j(t)|) \tag{3a}$$

$$G_{\alpha\beta}^s(r,t) = \frac{V}{4\pi N_\alpha N_\beta r^2} \sum_{i\in\{\alpha\}} \delta(r - |r_i(0) - r_i(t)|) \tag{3b}$$

Physically, $G_{\alpha\beta}^d(r,t)$ is the probability density of finding a particle $j$ at a distance of $r$ at time $t$ from a different particle $i$ at the origin $r = 0$ at time 0, therefore reporting on the correlations between different ions. Meanwhile, $G_{\alpha\beta}^s(r,t)$ is the probability density of finding a particle $i$ at time $t$ at distance $r$ away from its initial position at time $t = 0$, therefore reporting on the self-diffusion and transport properties of ions in the melt. The distinct parts corresponding to Figure 1 are shown in Figure S3(b)-(g), from which it is clear that $G_{\alpha\beta}^s(r,t)$ appears near $r = 0$ and is predominant up to 1.5 Å, while $G_{\alpha\beta}^d(r,t)$ is predominant at larger distances. By explicitly computing these components, interionic motions are clearly separated, enabling precise analysis of timescales, mechanisms, and transport properties.

Revisiting the total VHF in Figure 1(a), the positive correlation peak decay characteristic at 2.4 Å can now be seen as arising from a combination of simultaneous decorrelation between $Mg^{2+}$ and $Cl^-$ ions, increasing correlation between like ions (the distinct part of Mg-Mg and Cl-Cl VHFs), and self-motion of $Mg^{2+}$ and $Cl^-$ ions (the self-part of Mg-Mg and Cl-Cl VHFs) as shown in Figure 1(c)-(e) and S3 (c)-(g). While ion self-motion (the self-part of Mg-Mg and Cl-Cl VHF) is negligible at the 3.8 Å peak in the total VHF, there are still overlapping contributions from the simultaneous decorrelation of $Mg^{2+}$ to other $Mg^{2+}$ ions, and $Cl^-$ to other $Cl^-$ ions (Figure 1(d)-(e) and Figure S3(d)-(g)). As such, it is clear that the different dynamics cannot be readily inferred from only the total VHF spectra. Enabled by simulation, these separate dynamics are extracted and interpreted here.

Quantitatively, the temporal change in correlation between ions can be examined by analyzing the change in the distinct peak area of the partial (element-specific) VHFs. The peak area for positive correlation ($G^d_{\alpha\beta}(r,t) > 1$) is calculated in Equation 4.[8]

$$A(t) = \int_{R'_1}^{R''_1} [G^d_{\alpha\beta}(r,t) - 1]\, dr \tag{4}$$

Where $R'_1$ and $R''_1$ indicates the left and right minima of the first peak for $G^d_{\alpha\beta} > 1$. Figure 2 shows the correlation decay of Mg-Mg, Cl-Cl, and Mg-Cl ion-pair at 990 K, 1148 K, and 1600 K.

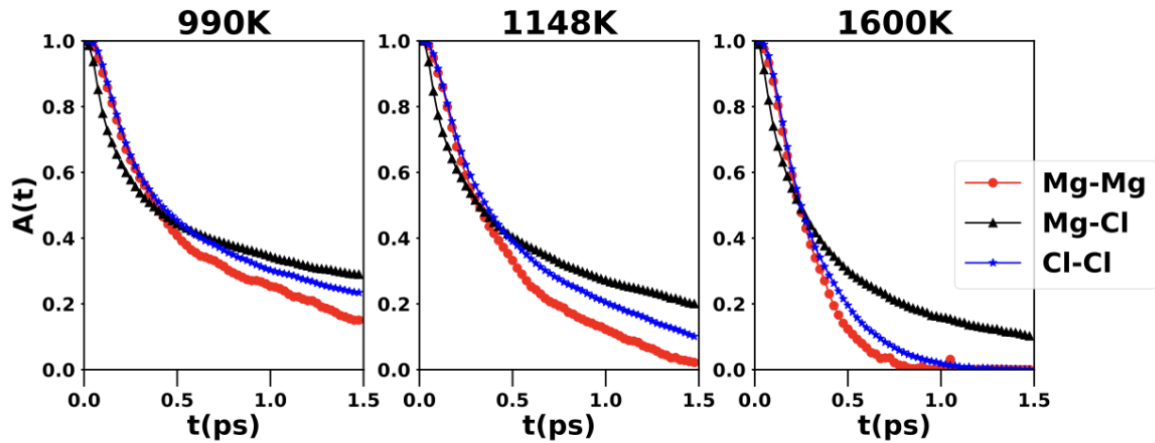

**Figure 2**. Positive correlation area decay with time for different ion-ion pairs in $MgCl_2$ simulated at 990 K (Left), 1148 K (Middle), and 1600 K (Right). At all temperatures, Mg-Mg decorrelation is the fastest, followed by Cl-Cl and Mg-Cl.

Here, a two-step decay process is observed, and $A(t)$ can be fit to a bi-exponential:

$$A(t) = A_1 e^{-\left(\frac{t}{\tau_1}\right)^{\gamma_1}} + A_2 e^{-\left(\frac{t}{\tau_2}\right)}, \tag{5}$$

where the time constants $\tau_1$ and $\tau_2$ represent the relaxation time for the first and second-step decay motivated by previous uses in capturing two-step relaxation dynamics in other liquids.[15] Here, the compressed exponential is used to encode multiple rapid and simultaneous processes, while the simple exponential describes the slower decorrelation of ions. Equation 5 is used for all pair-temperature combinations except for Mg-Mg at 1600 K, where it was found that such rapid decay processes could not be readily separated. In this case, only a single compressed exponential can be used to fit $A(t)$, due to the high temperature.

Physically, $\tau_1$ in Equation 5 is the time constant of a fast ballistic motion, while $\tau_2$ is the time constant associated with ion jumps across the first coordination shell.[14,15] $\tau_2$ values obtained from peak area decay for each partial VHF (Mg-Mg, Mg-Cl, Cl-Cl) are shown in Figure 3(a). First, the decay time for each ion pair decreases with increasing temperature. Secondly, a consistent trend is observed with $\tau_{2, \text{Mg-Cl}} > \tau_{2, \text{Cl-Cl}} > \tau_{2, \text{Mg-Mg}}$ across all temperatures. This trend confirms that ionic decorrelation is the slowest between oppositely charged ions Mg-Cl, followed by Cl-Cl and Mg-Mg, which can be rationalized by their respective ion-exchange energy barriers. This is given in terms of interaction potential ($w_{ij}^{eff}$) between each ion pair, as shown in Figure 3(b). The functional form of $w_{ij}^{eff}$ is given in Equation 6 and is written in terms of the partial RDF ($g_{ij}(r)$) and $\beta$, which is equal to $1/k_B T$.[20]

$$\beta w_{ij}^{eff} = -\ln(g_{ij}(r)) \tag{6}$$

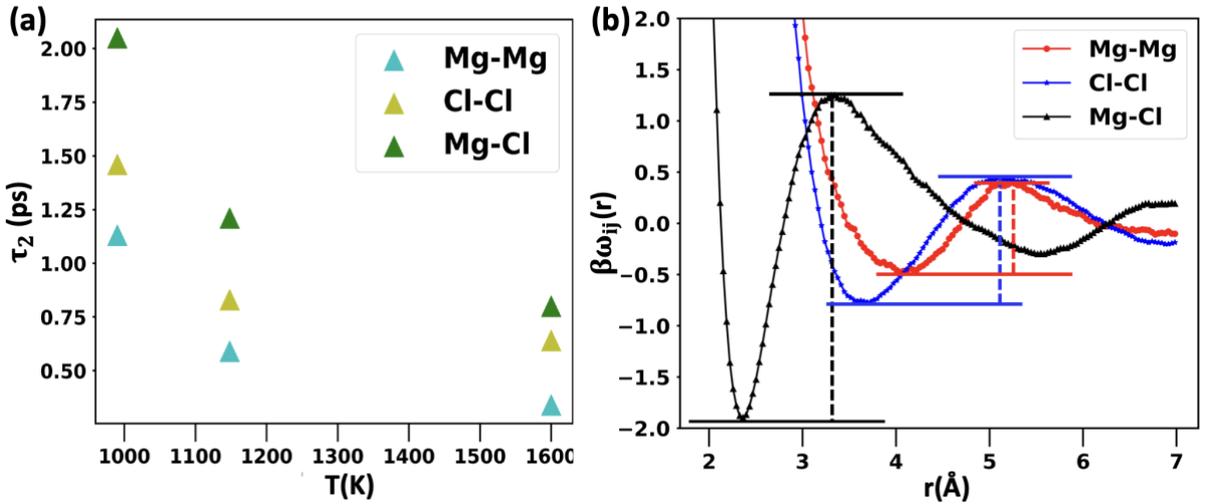

**Figure 3.** (a) Decorrelation time constants for Mg-Mg, Cl-Cl, and Mg-Cl at 990 K, 1148 K, and 1660K. For all temperatures, $\tau_{2, \text{Mg-Cl}} > \tau_{2, \text{Cl-Cl}} > \tau_{2, \text{Mg-Mg}}$. (b) Potential mean free energy of pair interactions in terms of $\beta \omega_{ij}$ (unitless). The barrier (dashed vertical line) for the Mg-Cl dissociation is the highest, followed by Cl-Cl and Mg-Mg.

The $\beta w_{ij}^{eff}$ as a function of the radial distance between each atomic pair at 1148 K is shown in Figure 3(b). Here, barrier heights, a.k.a. activation energies ($E_a^{\text{ion pair}}$) for ions to leave a solvation shell (difference between first maxima and minima of $\beta w_{ij}^{eff}$) are $E_a^{Mg-Cl} = 3.11\ \beta^{-1}$ (28.95 kJ/mol), $E_a^{Cl-Cl} = 1.2\ \beta^{-1}$ (11.6 kJ/mol) and $E_a^{Mg-Mg} = 0.87\ \beta^{-1}$ (8.3 kJ/mol). This shows that the association of opposite-charge pairs (Mg-Cl) is strongest, followed by Cl-Cl and Mg-Mg, which is aligned with their relative decorrelation times ($\tau_2$). In other words, the larger the activation energies for ionic dissociation, the longer the ion-ion decorrelation time, $\tau_2$, as one would expect. Herein, element-specific ion-ion correlations from atomistic simulations resolve ambiguities in the total VHF obtained from INS experimental measurements. Moreover, informing VHF with *ab initio* simulation enables insight into the atomic transfer mechanisms across the first solvation shell (SS), which is discussed further below.

The evolution of partial VHFs is found to coincide with Cl ion exchange between two [MgCl$_n$]$^{2-n}$ complexes, whereby this decreasing correlation between Mg-Cl in the SS is concurrent with the increasing correlation between Cl-Cl or Mg-Mg. This supports previous studies that suggested the existence of multiple co-existing coordination states of [MgCl$_n$]$^{2-n}$ that are hypothesized to frequently exchange between one another via corner or edge-shared chlorine atoms.[10,11,21,22] In other words, a chloride ion can dissociate from its original magnesium ion (decorrelation in the Mg-Cl VHF) by forming a metastable intermediate linked by Mg-Cl-Mg, which momentarily increases the correlation between Mg ions at $r = 2.4$ Å. Such atomistic mechanisms also explain decay behavior in the second positive peak at $r = 3.8$ Å, where the partial VHFs exhibit a decay in Mg-Mg/Cl-Cl correlation (separation of Mg ions following the transfer of a shared Cl from one Mg complex to another), along with an increase in Mg-Cl correlation (contraction of the Mg-Cl bond after the Cl transfer event). Figure S7 shows atomistic schematics of observed discrete ion transfer processes.

To further support the proposed ionic dissociation mechanisms, we compare VHF results with cage correlation function, $C_{in\text{-}out}(t)$.[23] While the cage correlation function cannot be measured experimentally, it is instructive as it explicitly quantifies decorrelation of the SS due to the transfer (inward or outward flux) of ions across the solvation shell and is calculated by Equation 7a-c:

$$C_{in-out} = \left\langle \theta\left(c - n_i^{out}(0,t)\right) \cdot \theta\left(c - n_i^{in}(0,t)\right) \right\rangle \quad (7a)$$

$$n_i^{out} = |l_i(0)|^2 - l_i(0) \cdot l_i(t) \quad (7b)$$

$$n_i^{in} = |l_i(t)|^2 - l_i(0) \cdot l_i(t), \quad (7c)$$

where $l_i$ is the nearest neighbor matrix, $\theta$ is a Heaviside function, and the $c$ is the number of atoms that must leave or enter an atom's neighbor list before it is considered that a change in surroundings has taken place. Here, $c = 1$ is defined to track single-atom exchanges within the SS. The timescales were obtained by fitting $C_{in-out}(t)$ with a bi-exponential function $ae^{-(t/\tau_1)} + be^{-(t/\tau_{LC})}$, where the $\tau_{LC}$ is the characteristic time scale for a single atom to leave or enter the SS. The cage correlation function for the Mg-Cl pair is shown in Figure 4(a), which shows an increasing decay rate with temperature as expected.

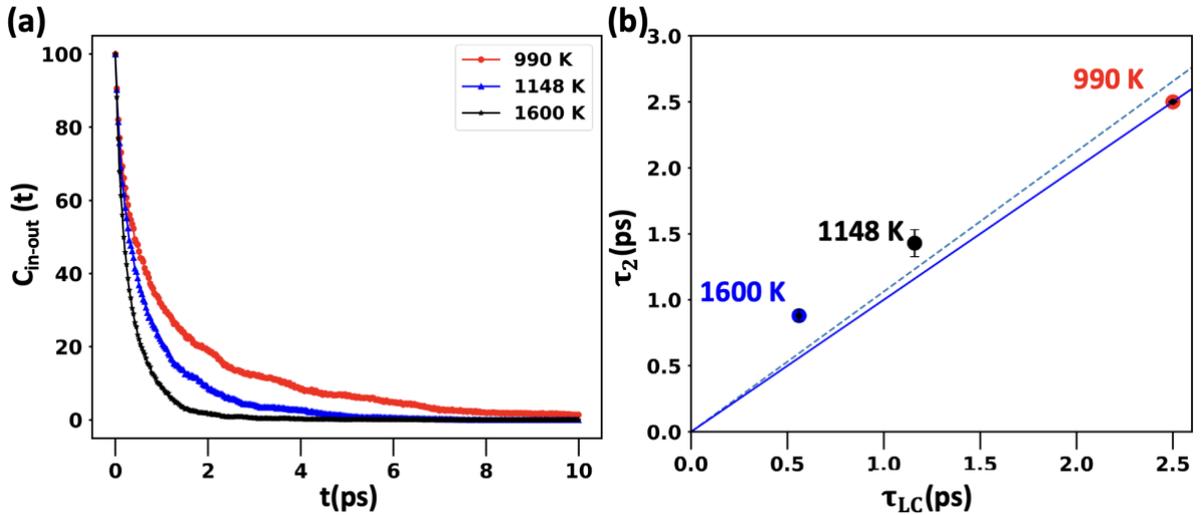

**Figure 4**. (a) Time-dependent cage correlation function $C_{in-out}(t)$ for Mg-Cl pairs at different temperatures. (b) VHF decorrelation time, $\tau_2$ versus the cage correlation time, $\tau_{LC}$ at 990 K, 1148 K and 1600 K. The solid line represents $C = 1$ in $\tau_2 = C\tau_{LC}$ and the dashed line is the least-squares regression line of the data in which $C = 1.14$.

The fitted $\tau_{LC}$ values from the cage correlation functions are compared with the $\tau_2$ values obtained from the analysis of the Mg-Cl partial VHFs in Figure 4(b). Here, $\tau_2 \cong 1.14\tau_{LC}$, suggesting that the VHF decay rate is closely related to single-atom exchange across the SS. Previous studies have shown that for $\tau_2 \cong C\tau_{LC}$[22], the coefficient $C$ depends on the relevant atomic stress relaxation mechanisms and bonding types. Namely, $C\sim 1$ was found for systems dominated by hydrogen bonding (e.g., water), whereas *C* ~ 4 was found for systems with metallic bonding (e.g., liquid metals like molten Fe).[8,24] Interestingly, our *C*-value of 1.14 would, therefore, suggest that ion dynamics in molten $MgCl_2$ are similar to those in water. In this case, an analogy may be drawn between the complex hydrogen-bonded molecular networks that have been found in water and the loosely-bonded cation-anion networks found in various molten salts, including $MgCl_2$.[25] Similarly, ion exchange of Cl between Mg ions discussed earlier points to an ion transport mode that is likely similar to the Grotthuss mechanism of hydrogen diffusion in water.[26,27] In fact, a Grotthuss-like mechanism has been previously proposed as the primary driver for electrical conductivity in $HgBr_2$ molten salt, whereby the charge is transported via Br-hopping between adjacent $HgBr_2$ molecules.[28] The finding of *C*~1 for $MgCl_2$ therefore suggests that a Grotthus-like transport mechanisms may exist beyond simple molecular liquids. Furthermore, previous works have suggested that $\tau_{LC}$ is very source of viscosity in high-temperature liquids, owing to its close relationship with the timescale for stress relaxation.[24,8] In fact, the viscosity of $MgCl_2$ (1.54 mPa-s at 1148 K) is similar to the viscosity of water at room temperature (1.01 cp at 293 K).[29,30] Likewise, the activation energy for viscous flow ($E_a$ in $\eta = Ae^{\left(\frac{E_a}{RT}\right)}$) in $MgCl_2$ is 21.97 kJ/mol, which is also similar to the activation energy for liquid water (15.7 kJ/mol) and to the activation energy for ion decorrelation determined by this study (8.3, 11.6 and 28.95 kJ/mol for Mg-Mg, Cl-Cl, and Mg-Cl at 1148 K respectively).[31,32] The results here therefore suggest a universal connection between local configurational dynamics and macroscopic transport properties across drastically different types of chemical environments (e.g., H-bonding in $H_2O$ versus ionic/covalent bonding in $MgCl_2$).

Herein, we lastly explore the ability of the VHF to directly recover ionic transport properties in molten salts, namely the self-diffusivity. To determine diffusion coefficients, a Gaussian function is used to fit the total self-part $G_s(r,t)$ (Figure S9), as shown in Equation 8.

$$G_s(r,t) = [2\pi\rho_1(t)]^{-3/2} exp(-r^2/2\rho_1(t)) \qquad (8)$$

Where $\rho_1(t)$ is a fitting parameter representing the mean-square displacement function.[14] Similar to water in previous work, the self-part of the VHF for molten salts is well described by the Gaussian approximation.[22,33] Using the $\rho_1(t)$ fit to the self-part of the VHF, the diffusion coefficient (*D*) is determined from the long-time fit of $\rho_1(t) = 2Dt$ shown in Figure 5(a). As shown in Figure 5(b), the diffusion coefficients increase with temperature, which corresponds to a decrease in relaxation timescale that was quantified previously by $\tau_2$ and $\tau_{LC}$.

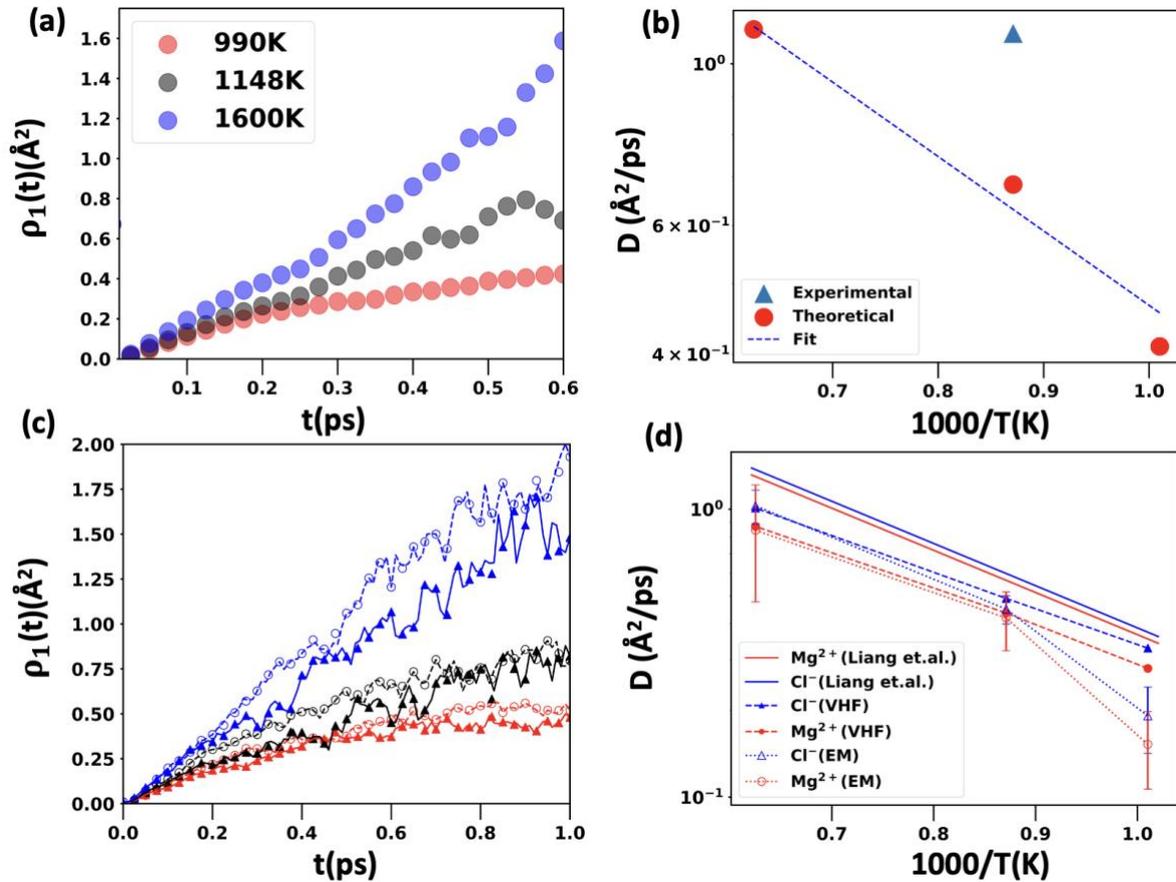

**Figure 5**. (a) Mean-square displacement $\rho_1(t)$ (or MSD in Å$^2$ unit) function obtained from fitting the total self-VHF $G_s(r,t)$. Uncertainty estimated by the standard deviation of the fitting was smaller than the size of the symbols. (b) Diffusion coefficients at 990 K, 1148 K and 1600 K from fitting $\rho_1(t) = 2Dt$. (c) MSD obtained from fitting partial element-specific self-VHFs for Cl$^-$ (open circle), and Mg$^{2+}$ (filled triangle) at 990 K (red), 1148 K (black), and 1600 K (blue). (d) Diffusion coefficients obtained from the NNMD (Liang et al.)[34], Einstein's block-averaging method (EM) (95% CI) and fitting the self-part of the partial VHFs.

Here, the total computed diffusion coefficient for MgCl$_2$ melt at 1148 K is ~0.7 Å$^2$/ps, which is slightly lower than the INS experimental reference diffusivity of 1.1 Å$^2$/ps at 1148 K.[35] One key reason for this difference could be the difference between the experimental and computational self-part. While the self-part can be extracted explicitly in simulation, it can only be approximated by the total

VHF in experiments, which breaks down at higher $r$ and $t$. However, it should be noted that provided the difficulties in measuring ionic self-diffusivity in high-temperature molten salts, a difference on the order of 0.4 Å²/ps (4x10⁻⁵ cm²/s) is not unreasonable. Namely, errors of similar magnitude can also be attributed to a combination of experimental and simulation uncertainties, such as controllable thermodynamic conditions, impurities, limited $Q$-range, finite energy resolution, and DFT approximations. The calculated diffusion activation barrier is determined by fitting the Arrhenius relationship $D(T) = D_0 \, exp\left(-\frac{E_a}{kT}\right)$ is 22.2 kJ/mol, which is comparable to the individual ion exchange barrier found by Figure 3(b). This suggests that diffusion in molten salt is governed by the exchange of atoms across SS, further supporting the idea of a Grotthus-like transport mechanism in $MgCl_2$. Recall, this is also similar to the activation barrier for viscous flow in both $MgCl_2$ (21.97 kJ/mol) and water (15.7 kJ/mol) providing further evidence to the idea of a universal connection between different fluids, and different macroscopic properties that can be traced back to common atomistic mechanisms/dynamics from which multiple properties emerge.

Beyond total diffusion, simulations allow us to extract ion-specific diffusion coefficients through the fitting of Equation 8 to partial self-VHFs, shown in Figure 5(c). These coefficients obtained through mean-squared displacement (MSD) fit to VHF, $D_{VHF}$, are compared to block-averaged diffusion coefficients calculated from the Einstein method $D_{EM}$, which uses the MSD of collective ion motions over long-time horizons explicitly computed from the AIMD trajectories.[2] Here, the self-diffusion coefficients reported by Liang et al. calculated from machine learning (ML)-based simulations are used as a benchmark, as shown in Figure 5(d). [34] This is chosen as a reference since large simulation cell sizes were enabled by highly accurate machine learning-based interatomic potentials,[36–39] avoiding known errors caused by intermediate-range ordering effects (and hence the calculated thermophysical properties) in AIMD cells.[38] Overall, the $D_{VHF}$ shows good agreement within 0.42 Å²/ps of the ML-based calculations for all temperatures, with similar magnitude of activation energies ($E_{VHF}^{Mg}, E_{VHF}^{Cl} =$ 24.1 kJ/mol eV and $E_{ML}^{Mg}, E_{ML}^{Cl} = 27.9$ kJ/mol). The $D_{VHF}$ and $D_{EM}$ match very well at the higher temperatures, but $D_{EM}$ underpredicts diffusion coefficient at 990 K, deviating from the Arrhenius relationship. The discrepancy is likely caused by the larger cell sizes and time horizons required for achieving a smooth EM-calculated MSD.[38] In contrast, VHF decorrelation time occurs on the picosecond timescale, allowing improved averaging with lower uncertainty, which is particularly noticeable at low temperatures. Overall, comparing all the results shown in Figure 5(d), we conclude that $D_{VHF}$ performs better than the commonly used $D_{EM}$, especially at lower temperatures, showing that VHF is a good tool for measuring diffusion coefficient. Interestingly, this suggests that macroscopically observed properties may be determined by interactions on microscopic scales as small as single digit Å in space and sub-picosecond in time, possibly down to the level of single ion exchange.

In summary, this work successfully unravels the element-specific dynamics behind the VHF and demonstrated the importance of a combined *ab initio*-INS experimental approach. Namely, ambiguities in interpreting the bond dissociation rates of different species from total INS spectra are resolved with well-validated *ab initio* simulations. While the simulated total VHF spectra agree well with experimental spectra at 1148 K,[14] it is clearly shown that competing correlations of different ion pairs and self-motions can obscure its analysis and preclude proper interpretation of standalone INS experiments. Using *ab initio*-informed VHF, relative decorrelation timescales $\tau_{Mg-Mg} < \tau_{Cl-Cl} < \tau_{Mg-Cl}$ were confirmed. Mechanistically, the partial VHF spectra are rationalized by the process of Cl ion exchange between $MgCl_n$ complexes and their fluctuations between various coordination states. This was supported by the calculation of the cage correlation function, which revealed a direct relationship between VHF decay and single ion exchange across the solvation shell. By comparing this relationship to other systems (*i.e.,* water), a universal connection between ion dynamics and macroscopic transport properties is proposed.[29,30] This is supported by similarities in the ion exchange mechanisms and activation energies of diffusion, bond dissociation, and viscosity that were common for both molten salt at high-temperature and water near ambient conditions. In doing so, VHF is also demonstrated as an accurate method for accessing and understanding transport properties. As such, our results support *ab initio*-informed VHF-INS as a tool that can significantly improve our understanding of the dynamics and chemical interactions of high-temperature melts and their connections to macroscopic properties.

**Methods**

AIMD simulations were performed using the Vienna Ab-Initio Simulation Package (VASP)[40] with the projector augmented wave (PAW) method, a plane-wave basis set, and the Perdew-Burke-Ernzerhof (PBE) generalized-gradient-approximation (GGA)[41] exchange-correlation functional. PAW-PBE pseudopotentials provided by VASP were used as Mg ($3s^2$) and Cl ($3s^2 3p^5$). A performance test has been conducted between pseudopotentials Mg ($3s^2$) and Mg_sv ($3s^2 3p^6$). Calculated RDFs from these two pseudopotentials match very well in terms of both peak intensity and peak positions (Figure S10). The computed average co-ordination number with these two different pseudopotential is also similar, , *i.e.* $n_{Cl-Cl}^{sim} = 11.6$, $n_{Mg-Cl}^{sim} = 4.4$, and $n_{Mg-Mg}^{sim} = 4.9$ using RDF cutoff of 3.45 Å, 5.1 Å, and 5.2 Å for Mg-Cl, Cl-Cl, and Mg-Mg, respectively. As such, the Mg pseudopotential was used to reduce computational expense. A large plane-wave cutoff of 650 eV with a 1e-5 eV convergence criterion for electronic self-consistent steps and a gamma-centered $1 \times 1 \times 1$ k-point mesh was used for reciprocal space sampling. The parameters are chosen to yield convergence within 2 meV/atom. The density functional theory (DFT)-D3 formulation proposed by Grimmes[42,43] was used to account for the effect of dispersion interactions. The initial structures were generated by randomizing the positions of the atoms (80 Cl and 40 Mg atoms) in a simulation box of a size consistent with the experimental density at 990 K, 1148 K, and 1600 K. The canonical ensemble (NVT)[44] using a Nosé-Hoover thermostat [45,46] was employed to sample the phase space at constant temperatures of 990 K, 1148 K, and 1600 K while maintaining the periodic boundary conditions. The integration time step of 1fs was used for all the calculations. All simulations were run for almost 50 ps, and the system was allowed to equilibrate for at least 12 ps before the structure and properties evaluation. The VHF was calculated with Python code based on MDtraj library[47].

**ACKNOWLEDGMENT**

We acknowledge the Department of Energy, Office of Nuclear Energy, Nuclear Energy University Program award no. DE-NE0009204 at the University of Massachusetts Lowell for financial support of this research. Work by Y.S. was supported by the U.S. Department of Energy (DOE), Office of Science, Basic Energy Sciences, Materials Sciences, and Engineering Division. Work by A.S.I. was supported as a part of the Molten Salts in Extreme Environments (MSEE) Energy Frontier Research Center, funded by the DOE Office of Science, Basic Energy Sciences.

**Supporting Information Available**:

Attached as PDF: Include information on VHF at different temperatures, self-part of VHF, distinct part of VHF, RDF, ADF, comparison of experimental and simulated VHF, and snapshot of ion transfer events.